# Computing support for advanced medical data analysis and imaging[*]

W. Wiślicki[1], T. Bednarski[2], P. Białas[2], E. Czerwiński[2], Ł. Kapłon[3],
A. Kochanowski[3], G. Korcyl[2], J. Kowal[2], P. Kowalski[1], T. Kozik[2], W. Krzemień[2],
M. Molenda[3], P. Moskal[2], S. Niedźwiecki[2], M. Pałka[2], M. Pawlik[2], L. Raczyński[1],
Z. Rudy[2], P. Salabura[2], N.G. Sharma[2], M. Silarski[2], A. Słomski[2], J. Smyrski[2],
A. Strzelecki[2], A. Wieczorek[2], M. Zieliński[2], N. Zoń[2]

**Abstract**
We discuss computing issues for data analysis and image reconstruction of PET-TOF medical scanner or other medical scanning devices producing large volumes of data. Service architecture based on the grid and cloud concepts for distributed processing is proposed and critically discussed.

**Introduction**

According to the World Health Organization's program of cancer control, more than 40% of cancer cases can be prevented, and even larger percentage can be cured if early detected [1]. To this end, one of the key practices to be implemented and made a medical routine is an early detection of small lesions and widespread monitoring of functionality of organs. Realization of these objectives requires that an accurate and frequent, non-invasive medical examinations are made available to wide public.

Persistent efforts in this direction are being continuously undertaken and are devoted to inventing still better resolution and cheaper medical scanning detectors. As an obvious consequence of detectors' refinement in precision and speed, dealing with data streams coming out of these equipments, or needed to simulate and optimize them, by far exceed earlier experience. This situation has already been observed earlier in high-energy physics and astrophysics, where new generation of detectors and data acquisition electronics was introduced one or two decades ago. Also in medicine it becomes clear that in order to handle efficiently growing streams of data, one has to equip both the detector developers and medical teams with fast data processing and digital image reconstruction methods and services.

In this paper we outline major computational problems related to data analysis and imaging. After formulation of these problems in terms of information technologies, we discuss some real-life solutions. In addition to its purely technical side, the problem has other aspects, as e.g. requirements of medical data protection. We outline the program of research and prospective service support in medical image processing for the novel solution of the Positron Emission Tomography (PET) based on the plastic Time of Flight (TOF) detector being developed at the Jagiellonian University [2]. This research and support program in the information technology domain is foreseen, both in the near and more distant future, as a major task for the Computing Centre CIŚ at the National Centre for Nuclear Research [3].

## Detector simulations, optimization and data reconstruction

In case of PET-TOF scanner, computing-intensive tasks previous to final imaging include simulations required for device design and optimization, scanner calibration, monitoring and event reconstruction. Computational and analysis framework J-PET was designed for these purposes [4]. It is based on the BOOST programming framework and general-purpose libraries written in C++ [5], Doxygen code documentation support [6], ROOT analysis and processing libraries [7]. Extensive simulations, accounting for detailed physical description of underlying processes of interaction of radiation and particles with detector materials, are performed using dedicated GATE system based on GEANT4 package [8].

## The basic problem and computing demands for digital imaging

Two steps are usually discerned in a standard approach to digital representation of physical objects: the *modelling* and *rendering*.

Modeling consists of development of mathematical representation of an object, be it numerical or closed-form formulas, of a 3-dimensional (3D) real volumes or 2D surfaces embedded in 3D space. Modeling may be technically very demanding process of gaining data from the real world, involving detection and data acquisition (DAQ) techniques. Real objects, being usually irregular and asymmetric, are almost always converted into non-parametric sets of volume elements (voxels). The first, raw modeling is performed by detector and DAQ system, providing raw data representation of an object. These sets are further reduced and used for image reconstruction. Their careful processing must give unbiased results and be checked by simulations in order not to destroy important details. Transferring data from DAQ to reconstruction processors, bandwidths of the order of Gigabits per second are normally sufficient. Since data are not stored on DAQ devices, transfers to servers next to them have to be very reliable in order not to lose data and therefore redundant links are needed.

The second step, rendering, is a technique of converting these sets of numbers or formulas, into a realistic picture, using methods of image reconstruction, simulating photo-realistic effects, combining 2D slices into complete 3D objects, smoothing, etc. Reconstructed data have to be kept in memory and are not further streamed. In real-life solutions, where time is a critical factor, rendering has to be a CPU-intensive processing, enabling strong data volume reduction (typically factor at least 5) and usually requiring concurrent processing: *threading, parallelization* or *vectorization*. The first one, threading, is a decomposition of a process into independent, executed in parallel subprocesses, and for codes foreseen to be executed on single processor it is normally done by compilers. This idea is implemented by switching the processor between threads located in different parts of memory, depending on the process phase. A more general concept, parallelization, may be implemented either on the single processor or many processors. In the first case, it is implemented as instruction-level parallelism and pipelining. At higher levels it may be realized as multi-core, distributed or grid computing. A special case of parallelism is represented by vector processing, being implementation of the *single instruction – multiple data* processing scheme.

One of the natural and most promising implementations of concurrent processing is given by the *Graphic Processing Units* (GPU) with many parallel execution units. Their advantages over CPU are also due to deeper pipelines (thousands instructions for GPU compared to 20 for typical CPU) and much faster memory interfaces, as they have to shift around more data. Typical GPU architecture is presented in Fig. 1, where basic stages of data transfer and picture rendering are presented. From the year 2000 on, GPUs tend to outperform traditional CPU servers for many applications where concurrent data processing may be efficiently used, image processing belonging to this class. Code development on GPU is not straightforward and requires familiarity with dedicated programming environments for heterogeneous computing platforms: *OpenCL* or *CUDA*.

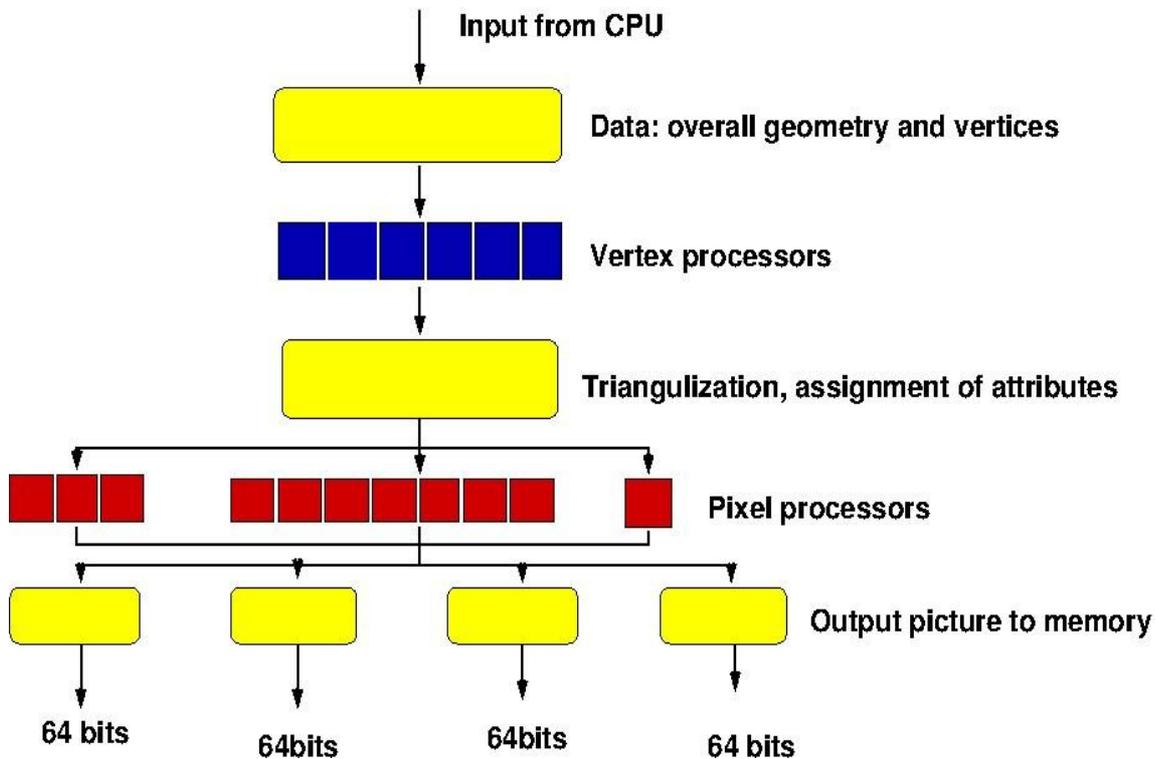

*Fig. 1: The GPU pipeline of data processing for image rendering, starting from data stream from CPU till final picture output from pixel processors (based on presentation by prof. Roger Crawfis, Ohio State Buckeyes).*

Considering an appropriate processing scheme for medical imaging applications, one has to decide first if it is going to be a localized or distributed computing. For a local model, either a usual CPU cluster or a mixed CPU-GPU server, or even a vector machine, obvious requirements of data discretion are usually easier to meet. But these resources have to be large and reliable if big sets of high-resolution pictures need to be obtained almost interactively. This requirement makes costs of an overall medical facility rapidly blowing up, sometimes in contrast to initial hope that these costs can be kept low. In addition, it is not going to be a one-time investment cost but a continuous exploitation burden which, in modern high-performance computing centers, even prevails investments in terms of money spending. Specifics of interactive applications, usually demanding resources in narrow time peaks, makes local computing facilities usually not flexible enough. All this puts more weight on distributed solutions.

## Solutions for distributed medical computing

Choosing distributed mode one deals with the following questions:

- Who owns and who administrates computing, memory and network resources?

Two solutions are seen for todays distributed computing networks. In the first one, all contributing parties agree between themselves and nobody waives ownership. In the second one, there is one owner and administrator. The first is cheaper, the second - more reliable.

- Who pays for infrastructure and services?

It may be either the resource provider who has to find finances outside of the user community, or just the users. The first case is rarely met unless public agencies or other wealthy parties support these network and service layers. In the second solution, the service is payable and thus costly for users, but they keep better control on its quality.

- Since medical data are touchy and in many countries are strictly protected by law, how data security and reliability is going to be ensured?

Basic security can be provided in the usual framework of the *Public Key Infrastructure,* extensively implemented and used nowadays, where transferred data are encoded using a public and private cryptographic key pair that is obtained and shared through a trusted authority. In this scheme, typical in academic and research non-secure networks, usually neither financial nor legal responsibility for data is assumed and providers work on the best effort basis. Though technically sufficient, this approach does not meet all detailed security requirements if not complemented by software requirements, its validation procedures, system quality and others [9]. For a certified medical data centers, high demand of reliability (>99.99%) and a predefined data cybersecurity with legal and financial responsibilities are specified.

Two architectural solutions for distributed medical computing, already proved to be efficient but still not implemented on a mass scale, are the g*rid computing* [10] and c*loud computing* [11].

On the grid, computing and memory resources may be scattered geographically and shared between owners. Resources are interconnected but not managed centrally. Users do not pay for using them. Task-to-resource matching is provided and optimized by one or more central services called Resource Brokers. Fundamental logical units on the grid are sets of processors and mass storage servers called, respectively, *Computing Elements* and *Storage Elements*, equipped with queue systems and managed in sites by their local authorities. Leading examples of large grids are the multipurpose, large scientific networks: the *Worldwide LHC Computing Grid* (LCG) [12] and the *Open Science Grid (OSG)* [13], run by organizations in the European Union and the United States, respectively, but involving both resource providers and users. Examples also exist for more specialized grids offering medical services [14]. The middleware and service layers on the grid are put on top of a robust and high-speed backbone networks, such as e.g. the GEANT network in Europe. User communities are logically organized in *Virtual Organizations* sharing resources located in many places. Medical applications, and in particular medical imaging and image exchange, were from the beginning among the most important medical services, e.g pharmacokinetics using contrast agent diffusion, radiotherapy planning using 3D simulations with GEANT4 and GATE, magnetic resonance image simulations, 3D volume reconstructions using large sets of radiological data, and others. Recently, grid computing in medicine has entered its commercial phase.

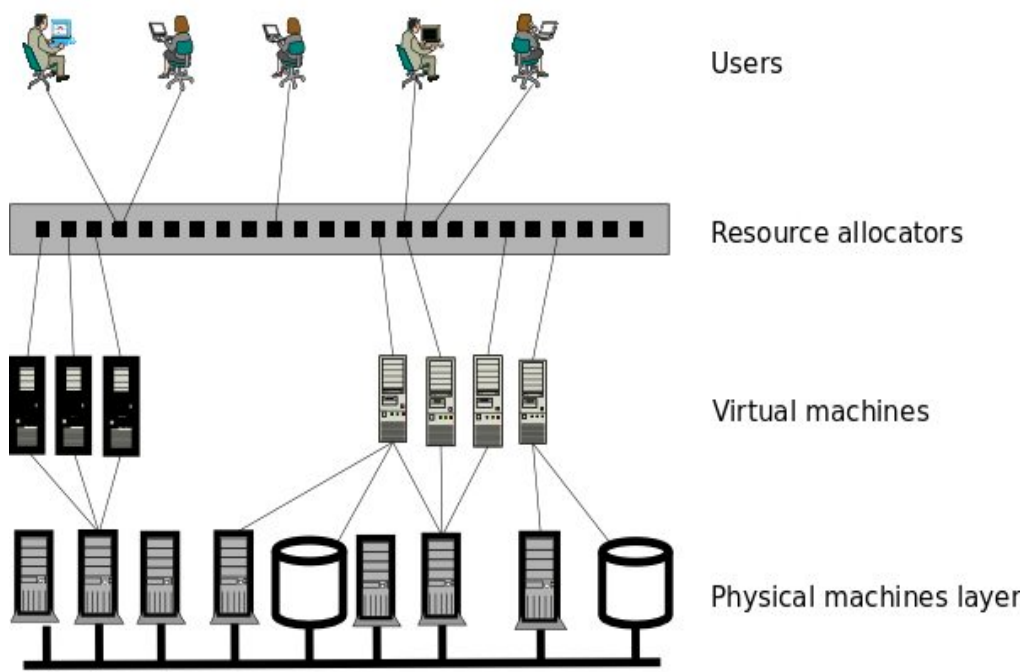

*Fig. 2: Scheme of the cloud service with characteristic four layers, in the bottom-top order: the physical fabric consisting of the computing and storage hardware, virtual machines being software computing units not identical to physical machines, resource allocator consisting of basic scheduling and brokering services, and the users.*

The general scheme of cloud computing is presented in Fig. 2. Its main advantage is smart virtualization that substantially optimizes usage of resources. Virtual machines are invoked specially for tasks and destroyed just after their completion, thus making disks and processors occupied when really needed. Tasks are matched to resources by using allocator services similar to resource brokers on the grid.

These two approaches to distributed computing, the grid and the cloud, were initially designed for different purposes and different user communities. However, in recent solutions computing elements on the grid tend to use typical cloud concepts. Remaining differences between these two, originally separate technologies, concern organization and proprietary issues.

## Workflow design for PET-TOF data processing

In Fig. 3 we present the workflow for fast image reconstruction on the grid and using computing resources deployed in the cloud scheme.

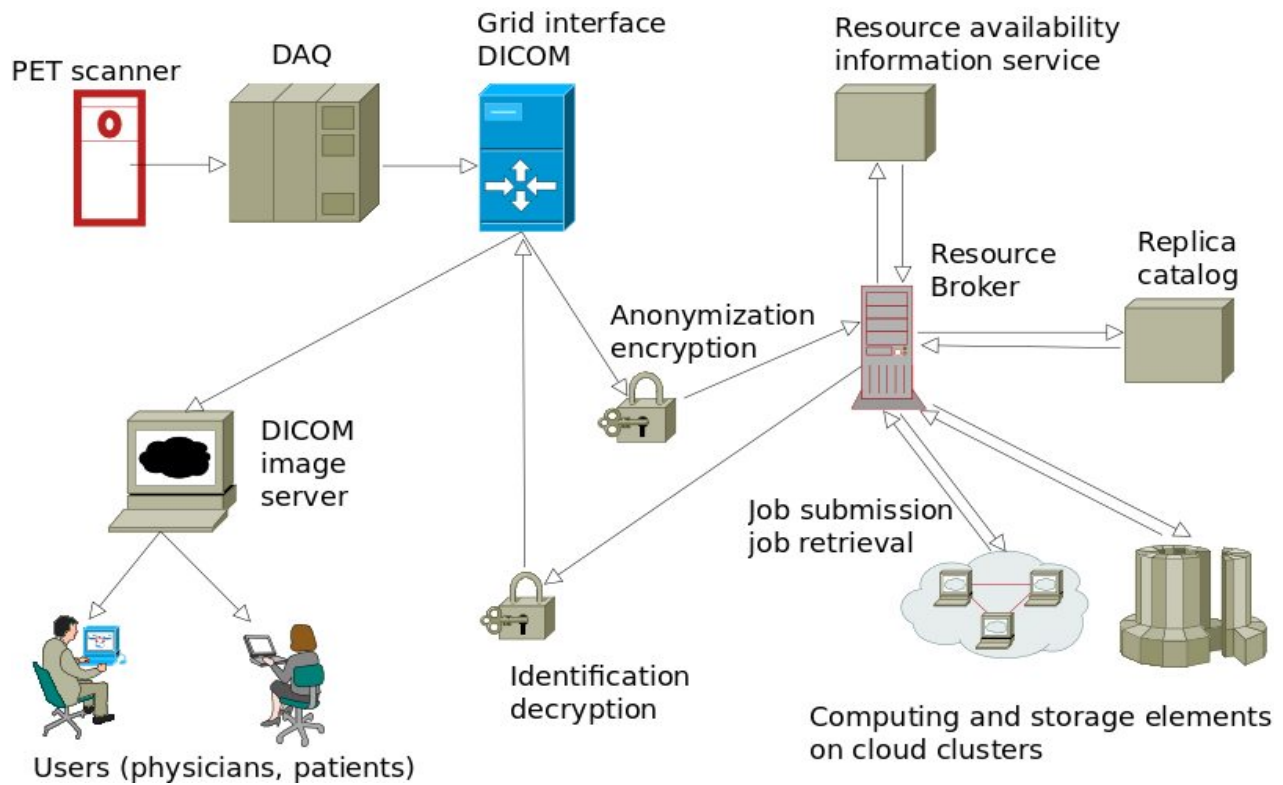

*Fig. 3: Workflow design for fast medical image reconstruction on the Grid.*

The DICOM services [15] are dedicated to medical image handling on the grid and are available commercially. Compared to many non-medical applications, medical data are sensitive and require special security treatment. Therefore, in addition to the usual security measures incorporated on the grid to any data propagated over the network outside of the scanner host site, real medical data need to be anonymized and encrypted, and identification keys be stored in secure memory. This functionality is normally provided by DICOM servers but due to its importance it is indicated separately in Fig. 3.

Successful merging of the original grid and cloud technologies ensures optimal resource usage. For the whole network it is given by resource brokers and in local computing elements and storage elements it is ensured by virtualization techniques. In order to work satisfactorily and provide with interactive image provision, the whole data exchange needs to have dedicated bandwidth secured at the level not less, and preferably higher, than 1 Gigabit per second and appropriate computing resources booked. Virtual machines are invoked by local gatekeeper machines when jobs are submitted.

# Conclusion

We outlined the concept of fast and highly efficient data analysis scheme intended to be used for processing data from the TOF-PET scanner. Most of these solutions are already known and partially implemented. We argue in favor of using distributed computing technologies in order to ensure the speed indispensable for interactive imaging. At the same time it suppresses unlimited blowup of local computing resources and costs entailed by that. Implementation of the idea of cheap and finely granulated PET scanner, being a necessary condition for breakthrough in the field of cancer diagnostics, is going to provide a flood of data of the order of tens of Gibabits per second or larger. It reminds the initial phase of modern, high-precision scientific measurements, and also contemporary everyday life, where enormous volume of detectors' outputs almost exceeds our capability to profit from them.

# Acknowledgements

This work was financially supported by the EU and MSHE Grant No. POIG.02.03.00-161 00-013/09, the Polish National Center for Development and Research through grant No. INNOTECH-K1/IN1/64/159174/NCBR/12, the Foundation for Polish Science through MPD Programme and Doctus - the Lesser Poland PhD Scholarship Fund.

# References

1. (2007). Cancer control: Knowledge into action. Early detection. Geneva, CH: World Health Organization
2. P. Moskal et al. (2012). Strip-PET: a novel detector concept for the TOF-PET scanner, Nucl. Med. Rev. 15 Suppl C, C68
   P. Moskal et al. (2012). TOF-PET detector concept based on organic scintillators, Nucl. Med. Rev. 15 Suppl C, C81
3. Świerk Computing Centre at the National Centre for Nuclear Research (2014). http://www.cis.gov.pl
4. W. Krzemień et al. (2013). J-PET analysis framework for the prototype TOF-PET detector, sumitted to Bio-Algorithms and Med-Systems, http://arXiv.org/abs/1311.6153
5. (2013). Boost C++ Libraries. http://www.boost.org
6. (2013). Doxygen. http://www.doxygen.org
7. R. Brun and F. Rademaker (1997). Root – An object oriented data analysis framework, Nucl. Instrum. Meth. A 389(1-2), 81-86. DOI: http://dx.doi.org/10.1016/S0168-9002(97)00048-X, http://root.cern.ch
8. S. Jan et al. (2004). GATE: a simulation toolkit for PET and SPECT, Phys. Med. Biol. 49(19), 4543-4562, DOI: http://dx.doi.org/10.1088/0031-9155/49/19/007, http://www.opengatecollaboration.org
   S. Agostinelli et al. (2003). GEANT4: A simulation toolkit, Nuclear Instrum. Meth. A

506(3), 250-303. DOI: http://dx.doi.org/10.1016/S0168-9002(03)01368-8, http://geant4.web.cern.ch
9. (2005). Guidance for Industry: Cybersecurity and Networked Medical Devices. Washington DC, USA: US Food and Drug Administration. http://www.fda.gov/MedicalDevices/DeviceRegulationandGuidance/GuidanceDocuments/ucm077812.htm
10. B. Jacob et al. (2005). Introduction to Grid Computing. http://ibm.com.refbooks
    I. Foster and K. Kesselman (2004). The Grid 2: Blueprint for a New Computing Infrastructure. Amsterdam, NL: Elsevier
11. I.Ivanov, M. van Sinderen and B. Shishkov (2012). Cloud Computing and Services Science. Berlin, D: Springer
12. (2014). Worldwide LHC Computing Grid. http://wlcg.web.cern.ch
13. (2014). The Open Science Grid. http://www.opensciencegrid.org
14. S. Benkner et al. (2005). GEMSS: grid infrastructure for medical image provision, Methods Inf. Med. 44(2), 177-181
    M.Garcia Ruiz et al. (2011). MantisGRID: a grid platform for DICOM medical images management, J. Dig. Imaging 24(2), 271-283, DOI: http://dx.doi.org/10.1007/S10278-009-9265-X
15. (2014). The DICOM Grid. http://www.dicomgrid.com